# Ionic Control over Ferroelectricity in 2D Layered van der Waals Capacitors

Sabine M. Neumayer[1], Mengwei Si[2], Junkang Li[2], Pai-Ying Liao[2], Lei Tao[3,4], Andrew O'Hara[3], Sokrates T. Pantelides[3,5], Peide D. Ye[2], Petro Maksymovych[1], Nina Balke[1,6*]

[1] Center for Nanophase Materials Sciences, Oak Ridge National Laboratory, Oak Ridge, 37831 TN, USA
E-mail: balken@ornl.gov

[2] Birck Nanotechnology Center and School of Electrical and Computer Engineering, Purdue University, West Lafayette, Indiana 47907, USA

[3] Department of Physics and Astronomy, Vanderbilt University, Nashville, Tennessee 37235, USA

[4] University of Chinese Academy of Sciences & Institute of Physics, Chinese Academy of Sciences, Beijing 100190, China

[5] Department of Electrical and Computer Engineering, Vanderbilt University, Nashville, Tennessee 37235, USA

[6] Department of Materials Science and Engineering, North Carolina State University, Raleigh NC 27695-7907, USA

*Corresponding author:
Nina Balke, ninabalke@ncsu.edu

**ABSTRACT**

*The van der Waals layered material $CuInP_2S_6$ features interesting functional behavior, including the existence of four uniaxial polarization states, polarization reversal against the electric field through Cu ion migration, a negative-capacitance regime, and reversible extraction of Cu ions. At the heart of these characteristics lies the high mobility of Cu ions, which also determines the spontaneous polarization. Therefore, Cu migration across the lattice results in unusual ferroelectric behavior. Here, we demonstrate how the interplay of polar and ionic properties provides a path to ionically controlled ferroelectric behavior, achieved by applying selected DC voltage pulses and subsequently probing ferroelectric* *switching during fast triangular voltage sweeps. Using current measurements and theoretical calculations, we observe that increasing DC pulse duration results in higher ionic currents, the build-up of an internal electric field that shifts polarization loops, and an increase in total switchable polarization by ~50% due to the existence of a high polarization phase which is stabilized by the internal electric field. Apart from tuning ferroelectric behavior by selected square pulses, hysteretic polarization switching can even be entirely deactivated and reactivated, resulting in three-state systems where polarization switching is either inhibited or can be performed in two different directions.*

1. INTRODUCTION

The multifunctionality of van der Waals (vdW) layered copper indium thiophosphate $CuInP_2S_6$ (CIPS), including ferroelectricity,[1-5] negative electrostriction,[6-7] and ion conductivity,[8-10] provides interesting opportunities for nanoelectromechanical devices and information storage. The layered vdW structure[11-12] allows cleavage of the crystal into ultrathin flakes that are still electromechanically active down to tens of nm in thickness.[13-14] The cleaved surfaces of these quasi-two-dimensional (2D) materials provide clean interfaces without dangling bonds and are therefore highly suitable for creating heterostructures with 2D materials such as $MoS_2$ or graphene to further enhance functionality, *e.g.* in ferroelectric field-effect transistors.[15] Below the Curie temperature of ~310 K, CIPS is ferrielectric with a polarization of typically 2.5 - 5 $\mu C/cm^2$ observed in hysteresis loops measured on micrometer scale electrodes.[1, 16] Although the polarization is lower than for other ferroelectrics, such as perovskite oxides, a relatively high electrostrictive coefficient leads to comparable longitudinal piezoelectric output.[6-7]

It was recently discovered that the polar properties of CIPS are more complex on the nanoscale. A combination of density functional theory (DFT) calculations and local piezoresponse-force-microscopy (PFM) discovered that there are actually two distinct iso-symmetric polar phases present in CIPS with a theoretical polarization of ~5 $\mu C/cm^2$ ( low-polarization phase, LP) and ~11 $\mu C/cm^2$ (high-polarization phase, HP), in which the Cu ions reside just inside or just outside of the layers, respectively.[4] As a consequence, the energy potential well as a function of Cu displacement shows a quadruple well instead of the more common double potential well exhibited by other ferroelectric materials. Copper-ion conduction is characterized by an activation energy of 0.73 eV from macroscopic characterization[8] and 0.71 eV from local characterization,[9] respectively. This is in good agreement with theory, which shows that the energy barrier for Cu-ion jumps across the vdW gap was calculated to be 0.65 - 0.75 eV in the presence of Cu Frenkel pairs and slightly higher in stoichiometric regions.[17]

As a material that exhibits both ferroelectricity and ionic conductivity, CIPS has the not-so-common property that the same atomic species, Cu, is responsible for both the short-range displacements that determine the polarization in the ordered, ferroelectric state below $T_c$[1, 4] and the long-range ionic current across layers and van der Waals gaps,[8-9] which is viewed as a disordered state. In 1999, Scott pointed out that in materials that feature such a property, at temperatures below $T_c$, voltage can be used to transition from the ordered ferroelectric state to the disordered state of ionic conductivity using $Ag_{26}I_{18}W_4O_{16}$ where Ag ions are the mobile species.[18] More specifically, he suggested that for small voltages, the polar atomic displacements can be flipped back and forth within the double-well potential by reversing the field, while at large voltages above a threshold value the energy barrier for displacements between unit cells would be overcome and the ions would move without bound. Therefore, a strong dependence of ferroelectric material properties on voltage amplitude and duration is expected. Indeed, it has been shown that the frequency of electric-field cycles affects polarization switching in CIPS. Most notably, the polarization increases with decreasing frequency of the electric field cycle and the hysteresis loops appear imprinted for high frequency field cycles.[10] This behavior has been explained by the rotation of ionic defect dipoles rather than by ionic conductivity.[10] However, it has already been demonstrated that at room temperature, *i.e.*, in the ferroelectric phase, application of voltage pulses can drive the Cu atoms across the vdW gap, which leads to highly unusual polarization reversal processes such as polarization alignment against the applied electric field. This phenomenon has been demonstrated for locally biased measurements with an atomic force microscope (AFM) probe across one vdW gap[19] as well as under biased electrodes across multiple vdW gaps.[17] These observations demonstrate that it is possible to explore the regime of coexisting ferroelectric and ionic properties but it has not been leveraged yet to actively control material functionality, e.g., ferroelectric properties.

In this work, we activate ionic conductivity at room temperature using electric fields and study its impact on ferroelectric switching in CIPS micro-capacitors. After activation, we found that polar and ionic properties co-exist and contribute this to measured currents. Based on current-voltage characteristics we can distinguish between ionic and polarization-switching currents and study the symbiotic interplay of polarization and ionic conductivity. We find that the presence of increasing ionic currents alters the timing and intensity of switching peaks, which in turn leads to evolving polarization loops, demonstrating direct control of polarization loops by ionic currents. These changes in the polarization hysteresis can be explained by DFT calculations in terms of an internal electric field that builds up as ionic currents drive an increasing number of Cu ions to the top CIPS-electrode interface, while an increasing number of CIPS layers at the bottom of the stack are depleted of Cu ions and become metallic, maintaining the circuit integrity.

Moreover, the internal field favors the HP state over the LP state in the direction of the electric field, which explains the increase in measured polarization. In the extreme case at higher voltages, ionic conduction was found to entirely suppress ferroelectric hysteretic switching, a process that is reversible, allowing turning the ferroelectric characteristic on and off.

## 2. RESULTS AND DISCUSSION

### 2.1 Measured Relationship between Ionic Currents and Ferroelectric Behavior

Square and triangular voltage, $V$, pulses are applied to the top electrode of a ferroelectric CIPS capacitor (see schematic in Figure 1(a)) via micromanipulators to activate ionic currents and study their influence on the switching behavior in subsequent fast triangular voltage sweeps. In a first step, we applied a square pulse of varying duration followed by triangular $V$ sweeps on a pristine CIPS capacitor and recorded the resulting currents, $I$. Figure 1(b) shows the voltage, $V$, waveform consisting of DC pulses of 4, 10, 20, 100, 200, 400, 500, 600, 800, or 900 ms duration at an amplitude of -4 V. Subsequently, the ferroelectric behavior was probed by two triangular $V$ sweeps of 10 ms duration each, as typically performed for ferroelectric characterization. The amplitudes of the DC pulse and the triangular $V$ sweeps of 4 V were chosen to clearly exceed the coercive voltages that lead to ferroelectric switching in this ~185 nm thick sample. Figure 1(c) shows the currents measured during the entire duration of all DC voltage pulses. For each DC voltage pulse, a polarization switching current peak is detected within the first 2 ms. For voltage pulses longer than 100 ms, significant current is detected, which grows exponentially to exceed the specific current measurements limits of -256 nA for pulses longer than 400 ms. This second current contribution is assigned to an ionic current indicating Cu ions move beyond the unit cell. It is evident that the DC voltage step can be used to control the amount of ionic current. The currents at the beginning and end of each DC voltage pulse are shown in detail in Figure S1. Figure 1(d) depicts the currents during the subsequent two triangular voltage sweeps at 100 Hz. Up to 100 ms duration of the preceding DC voltage pulse, only polarization switching current peaks corresponding to ferroelectric displacements are observed. There are no switching peaks at the negative voltage in the first cycle as the system is already switched to a positive polarization orientation due to the negative polarity of the DC voltage pulse. This feature agrees with standard ferroelectric behavior. For longer DC voltage pulses, additional current contributions for $V < 0$ become visible, indicating ionic current contributions during 100 Hz field cycling. However, the current peak assigned to polarization switching for $V < 0$ is still visible. It shall be noted that the situation reverses for positive DC pulses that lead to positive ionic currents during the DC pulse and the triangular $V$ sweep (Figure S2). The polarity of the DC pulse determines the direction of the ionic current observed in subsequent fast triangular voltage sweeps. Since ion transport is activated in the direction of the applied electric field, Cu accumulation occurs at the electrode that is negatively biased. Therefore, the charge distribution is asymmetric during the triangular voltage sweeps, which manifests as an asymmetry in ionic current dependent on voltage polarity. The ionic nature of the measured large current under negative (positive) DC pulses and the negative (positive) portion of the triangular voltage sweep is confirmed by the exponential increase with time or voltage amplitude. Moreover, electronic leakage typically erodes polarization switching which still occurs in our measurements. The total current, $I$, can be divided into two types corresponding to their origin: (i) ionic currents proportional to the applied voltage for negative voltage polarity and (ii) switching current peaks for positive and negative voltage polarity originating from charge redistribution at the electrodes after polarization reversal. Separating these current contributions allows the study of ferroelectric switching in the presence of ionic currents.

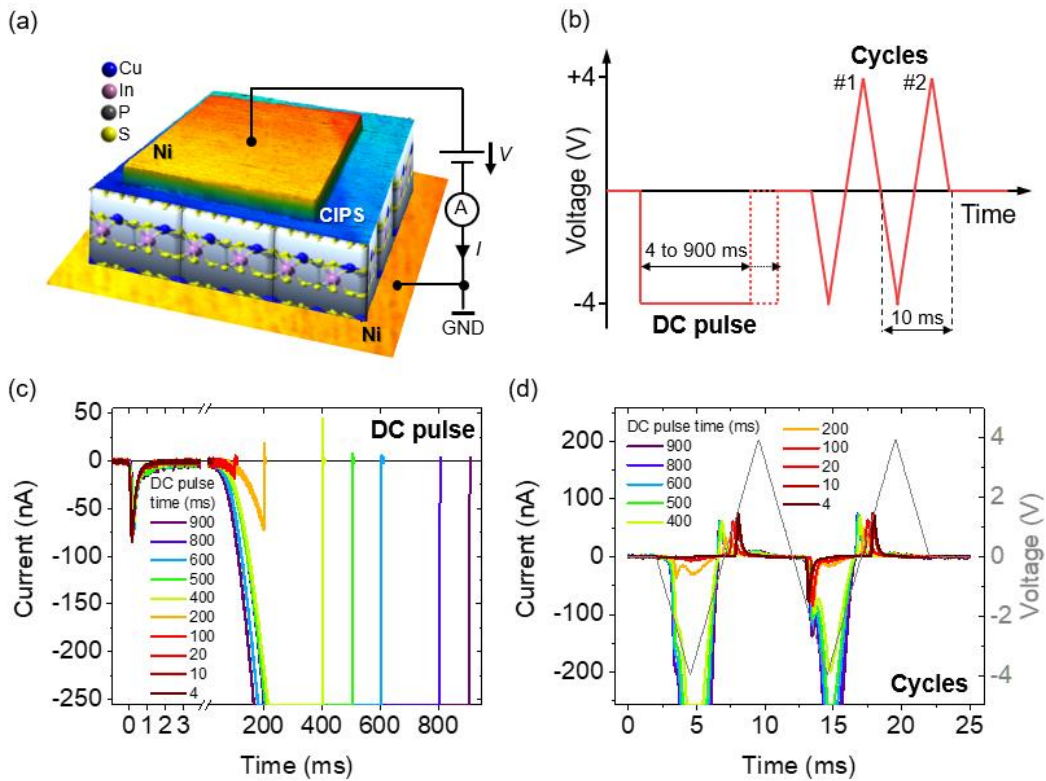

**Figure 1.** (a) Schematic depiction of experimental setup and (b) applied voltage sequence showing -4 V DC voltage pulses with a duration of 4 to 900 ms (10 steps) preceding two triangular voltage cycles between ±4 V of 10 ms duration each. (c) Currents measured during DC pulses of varying duration. The square voltage pulses were turned on at 0 s. (d) Currents measured during triangular cycles. Labels in legends refer to DC pulse duration in ms.

In the following, we focus on the analysis of the current signal during the triangular voltage sweeps. As shown in Figure 1(d), ferroelectric and ionic current contributions are visible. To separate these contributions, a four-peak fitting method was applied to separate the two polarization switching peaks and the ionic contribution during the second triangular voltage sweep. The four-peak fitting method accurately captures the overall current behavior consisting of two peaks associated with polarization switching and two peaks representing ionic currents associated with ion transport and electrochemical processes in forward and reverse direction. The second cycle was chosen for the analysis since the first cycle did not describe the full polarization hysteresis loops. However, it is visible in Figure 1(d) that the ionic current contribution is reduced between the first and second cycle, which we will address later. The fitted ionic current contribution as well as the switching current peaks after the current separation are shown in Figure 2(a) and 2(b), respectively. The fits are shown in detail in Figure S3. The corresponding hysteresis loops of polarization obtained by integrating the polarization switching current peaks in Figure 2(c) are calculated via $P = -\frac{1}{2A}\int I(t)dt$, where $t$ is the time and $A$ is the area of the electrode. Note that the currents measured at the electrode occur in response to screening the ferroelectric polarization, where a negative current peak corresponds to ferroelectric switching for a more positive polarization state. This is an accepted procedure since only relative polarization changes are measured.

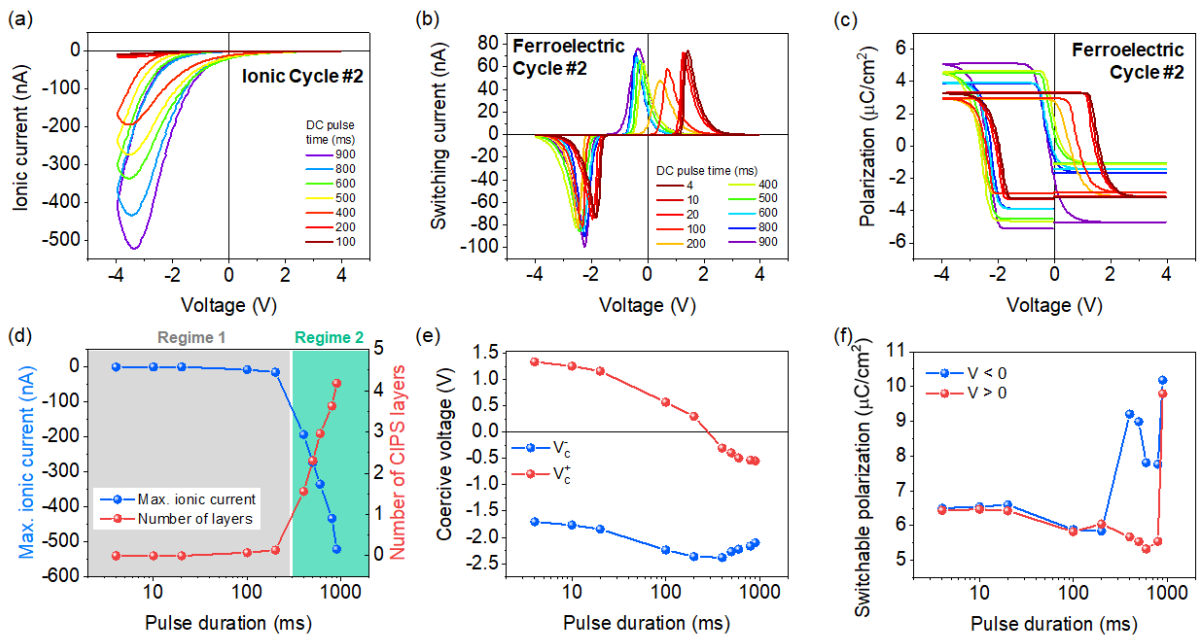

**Figure 2.** (a) Ionic current during triangular sweeps for the second cycle extracted from total current fitting. (b) Switching current calculated from total current fitting. (c) Polarization hysteresis loops calculated from (b). (d) Maximum ionic current as a function of DC voltage pulse duration and corresponding number of CIPS layers below the electrode area contributing to the ionic charge. Gray and green backgrounds indicate two regimes with significantly different amounts of ionic current. (e) Coercive voltages extracted from current peaks in panel (b). (f) Absolute value of switchable polarization on the left ($V < 0$) and right ($V > 0$) side of the hysteresis loop as a function of pulse duration.

It can be seen that the ferroelectric polarization hysteresis loop is affected by the ionic current. We can identify two regimes determined by the maximum ionic current and ionic charge that can be related to the number of CIPS layers that are needed to provide their Cu ions for electrochemical reactions so that the measured maximum ionic current can be reached (Figure 2(d)). Details on how the ionic current is related to ionic charge and number of CIPS layers under the electrode area and the provided the charge are shown in the Supplemental Information (Figure S4). Two clear regimes with different behavior are highlighted in Figure 2(d) by shaded backgrounds. The transition between the two different regimes coincides with the amount of charge the Cu from a single layer of CIPS can provide, 0.20 nC, under the assumption that each Cu ion reacts with one electron, as discussed below. In regime 1 up to 200 ms DC voltage duration, no (4 ms, 10 ms, 20 ms) or only minimal (100 ms, 200 ms) ionic currents are detected. In this regime, both coercive voltages shift toward negative voltages indicating the formation of an internal bias field that stabilizes the negative polarization direction (Figure 2(e)). The direction of the internal bias field is consistent with a positive charge accumulation at the top electrode, which is caused by an accumulation of Cu ions under the influence of the negative DC voltage pulses. The coercive voltage shift is consistent with observations by Zhou et al.[10] At the same time, the amount of switchable polarization with positive and negative voltages are identical and decrease slightly from 6.6 µC/cm$^2$ to 5.9 µC/cm$^2$ (Figure 2(f)). In regime 2, for DC pulse duration larger than 200 ms, the maximum ionic current during the triangular sweep increases linearly as a function of the logarithmic DC voltage pulse duration and grows >80 nA per 100 ms. In this regime, the positive coercive voltage continues to shift toward more negative voltages and becomes negative itself, while the negative coercive voltage reverses trend and shifts toward more positive voltages. This change is accompanied by asymmetric hysteresis loops (different amount of switchable polarization on each side of the hysteresis loop), which originate due to the increase of the negative polarization $P^-$. For the longest DC voltage duration of 900 ms, the positive polarization $P^+$ increases as well, making the polarization hysteresis more symmetric with respect to the polarization axis with an increased switchable polarization of about 10 µC/cm$^2$. The strong shift of both coercive voltages into the negative voltage regime leads effectively to a loss of remanent polarization reversal. It is also noteworthy that while the polarization hysteresis changes drastically, the area of the polarization loops, related to energy, is fairly constant and ranges between 0.24 J and 0.16 J (Figure S5).

Next, we discuss the observation of ionic currents during the 100 Hz voltage sweeps. Evidently, generating ionic current during >100 ms DC voltage pulses directly enhances ionic conductivity in the subsequent fast voltage sweeps for the same voltage polarity. We hypothesize that the ionic current is based on an electrochemical reaction of Cu ions at the electrode forming Cu metal. Based on the polarity of the DC voltage pulse, Cu ions accumulate under one of the electrodes (top or bottom) and partially react to Cu metal leading to current flow, $Cu^+ + e^- \rightarrow Cu$. In subsequent fast sweeps, ionic currents can be measured because accumulated $Cu^+$ is readily available at that electrode, which increases the kinetics of the electrochemical reaction. Over time, Cu redistributes, driven by a voltage-induced concentration gradient that minimizes the amount of $Cu^+$ accumulation under the electrode, leading to a smaller ionic current with each subsequent cycle as shown in Figure S6 for 10 cycles measured after DC voltage pulses of 20 ms, 500 ms and 1000 ms duration (Figure S7). Alternatively, the application of DC voltage pulses can lead to the formation of defects, such as Cu vacancies and Frenkel pairs, which lower the energy barrier to cross the vdW gap.[17] In this scenario, we have to assume that these defects are not stable over time or under continuous cycling.

**2.2 Theoretically Calculated Impact of Ion Accumulation and Depletion on Ferroelectric Hysteresis**

We now turn to the results of theoretical calculations based on DFT aimed to explore the connection between polarization hysteresis loops and ionic current in CIPS. We employed a stack of six CIPS layers with Ni electrodes (three layers of Ni atoms) using an in-plane unit cell that contains four formula units per CIPS layer. To simulate the effect of negative DC voltage pulses applied to the top electrode during the experiment, we start with a stochiometric CIPS stack where all Cu atoms are at the bottom of the layers in the -LP state (Figure 3(a)) and move the Cu sheets upwards across layers and vdW gaps, going through multiple negative and positive polarization states (Figure 3(b-h)). At each point, the system is fully relaxed. This process results in the formation of Cu-depleted CIPS layers at the bottom electrode and Cu accumulation underneath the top electrode, simulating Cu redistribution under externally applied fields. The number of empty CIPS layers vs. the number of fully stoichiometric CIPS layers in the stack are expressed by *m* and *n*, respectively, and the stack can be described by the general form of Ni/*m* InP$_2$S$_6$/*n* CuInP$_2$S$_6$/Cu/Ni (Figure 3(i)).

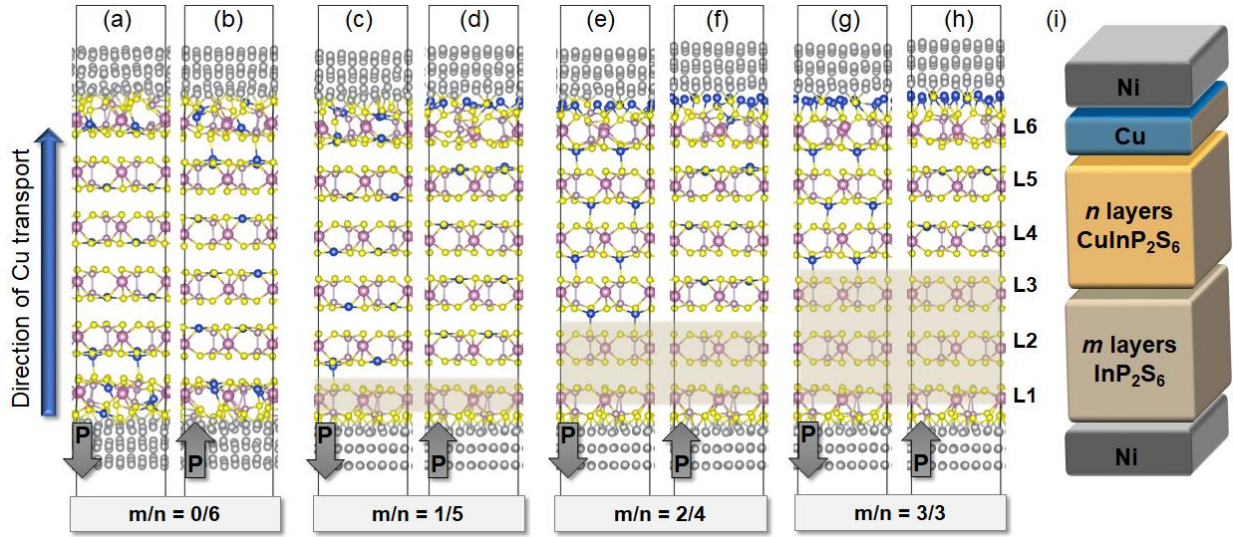

**Figure 3.** Stack of six CIPS layers with Ni-layer top and bottom electrodes at various stages of concerted, upward Cu motion. Relaxed structure of CIPS stack for (a,b) no empty CIPS layers ($m = 0$), (c,d) one empty CIPS layer ($m = 1$), (e,f) two empty CIPS layers ($m = 2$), and (g,h) three empty CIPS layers ($m = 3$) for the downward and upward polarized state, respectively. Cu-free CIPS layers are highlighted in beige. (i) General schematic of CIPS stack including Cu-empty and fully stochiometric CIPS layers and accumulated Cu under the top electrode for $m > 0$.

Noteworthy points are as follows. In the initial state, when all Cu atoms are in a negative polarization state, they occupy the -LP state, but Cu atoms in the top layer (L6) and the bottom two layers (L1/L2) undergo substantial relaxations (Figure 3(a)). When the Cu ions are moved across the layers to a positive polarization state (Figure 3(b)), they occupy the +LP state, but the relaxations now are substantial in the top two layers (L5/L6). In the next step (Figure 3(c)), the Cu sheets are moved across the vdW to negative-polarization states with the layer polarization aligned against the external electric field that in principle is driving the upward motion of the Cu ions. We now have an empty CIPS layer at the bottom and one Cu sheet under the top Ni electrode. The topmost Cu sheet does not spontaneously electroplate on the Ni electrode. Instead, S atoms tend to bond to Ni atoms across the interface. We checked that electroplating indeed does not take place by placing the Cu atoms on the Ni surface, but upon relaxation, Cu retreated to the nearest CIPS layer (a single CIPS Cu sheets corresponds to only ¼ of the Cu atoms needed to form a Cu monolayer on Ni). In the next step (Figure 3(d)), when we move the Cu sheets across the layers one more time, Cu atoms from two CIPS layers are at the top electrode interface and partial electroplating begins. In the next step, Figure 3(e), we move the Cu sheets across the gaps again, leaving two empty layers at the bottom. For the first time, Cu atoms in the "bulk" layers spontaneously acquire -HP positions, while in the subsequent step (Figure 3(f)), the relaxed Cu position is on the +LP polarization state, which is also true for the configuration with one more empty CIPS layer (Figure 3(g,h)).

In order to determine the metallic or insulating nature of the CIPS layers as we proceed through the evolution depicted in Figure 3, we calculated the projected densities of states (PDOS) for all the layers in each of the eight stacks. The most important finding is that the Cu-depleted CIPS layers at the bottom become metallic while the stoichiometric CIPS layers remain insulating. This result is shown for the example of two empty CIPS layers ($m/n = 2/4$) in Figure 4(a). The two bottom layers (L1/L2) clearly show metallic character while the layers containing Cu (L3-L6) maintain their insulating character. The PDOS for all configurations shown in Figure 3 are shown in Figure S8. We note that in a physical system with a very large number of CIPS layers, after emptying four CIPS layers at the bottom and electroplating a full monolayer of Cu on the top electrode, we effectively start again from the configuration of Figure 3(a) where we have a stack of stoichiometric CIPS layers between metallic electrodes, which are formed by Ni/InP$_2$S$_6$ at the bottom and Ni/Cu at the top. This result enables us to compare additional theoretical results obtained from the configurations in Figure 3 with the experimental data. We note that metallicity is not present in the Cu-free equilibrium phase of In$_{4/3}$P$_2$S$_6$ (Figure S9). Here, the metallic behavior arises due to the Cu-depleted layers in non-stoichiometric InP$_2$S$_6$.

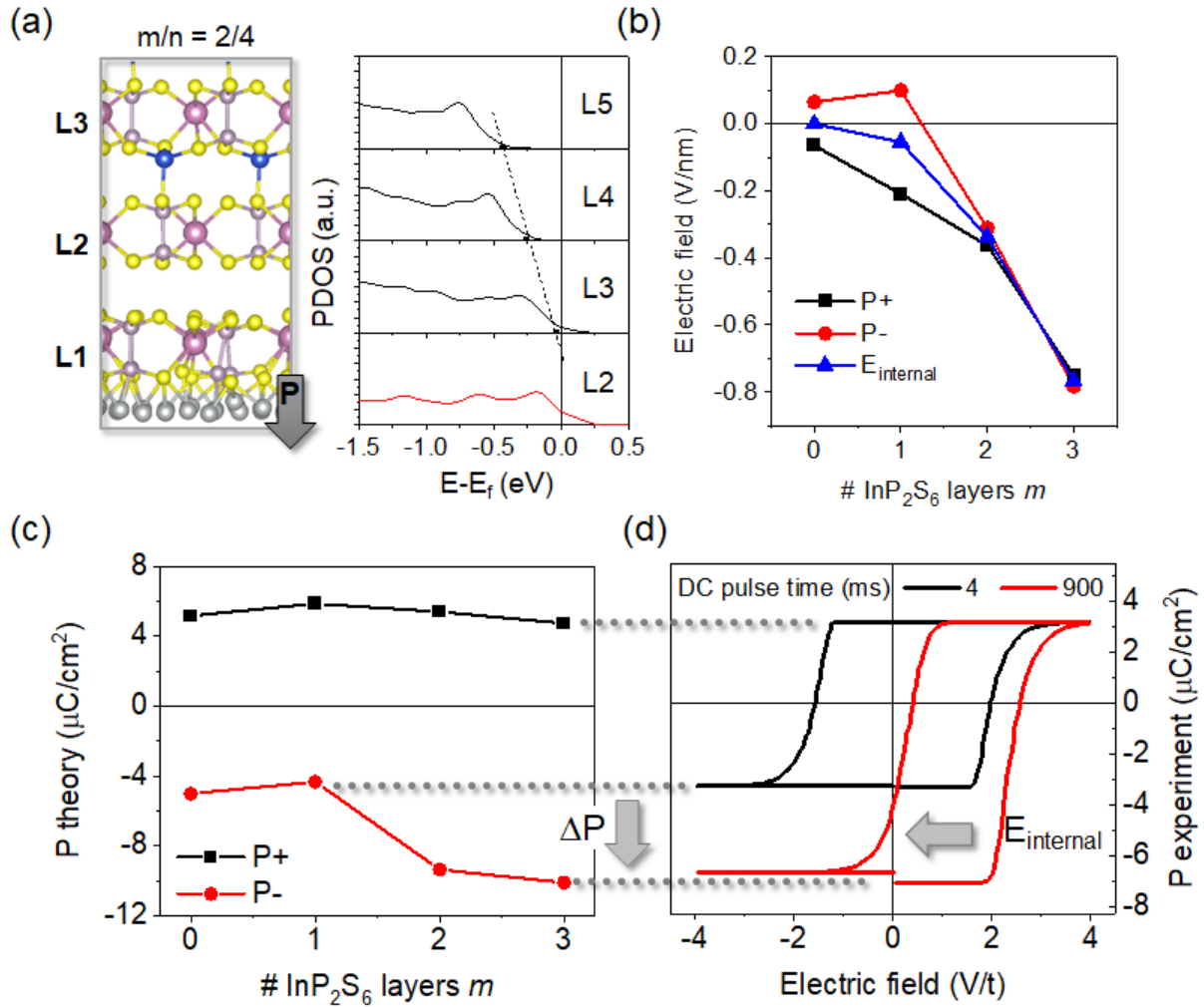

**Figure 4.** (a) Layer-projected DOS (PDOS) in the *m/n*=2/4 stack of Figure 3(e) showing metallic nature of Cu-free layer L2. Electrically insulating layers are drawn in black, conductive layers are drawn as red lines. Layer L1 and L6 are omitted as they are fully metallic due to contact doping in all cases. The slope of dashed line corresponds to value of internal electric field that arises from Cu migration to the top electrode. (b) Progression of calculated internal field for eight stacks of Figure 3 for each pair of polarization states and average of the two. (c) Calculated average polarization values per layer in eight stacks of Figure 3. (d) Experimental polarization loops for DC voltage pulses of 4 ms and 900 ms duration.

The PDOS plots also allow us to determine the total internal electric field, $E_{internal}$, that is generated by the movement of Cu atoms leaving empty layers at the bottom and placing Cu atoms at the top interface. We obtain the electric field in each stack by tracing the top of the valence bands through the insulating CIPS layers (dashed line in Figure 3(a); see also Figure S8). This trace is effectively the macroscopic electrostatic potential that is generated by the empty layers at the bottom and the accumulation of Cu atoms at the top interface. The slope of the linear trace is the corresponding $E_{internal}$. The potentials are shown in Figure S10(a) while the resulting electric fields as the system evolves are shown in Figure 4(b). The blue points correspond to the averaged $E_{internal}$, which is proportional to the number of empty CIPS layers, $m$. The sign of this field corresponds to pushing Cu atoms downward, *i.e.*, it opposes the external electric field that pushes Cu atoms upward.

Using the results obtained from the calculations for Figure 3, we derived the polarization values for each of the stacks. We used the calculated Cu displacements in each of the stoichiometric layers in the stack, averaging over the layers and then using the polarization-versus-Cu-displacement curve reported in Ref. [4] for bulk CIPS. The average polarization versus average Cu displacements are shown in Figure S10(b) and the extracted polarization values as functions of the number of empty CIPS layers, $m$, are shown in Figure 4(c). As described earlier, the positive +LP state is unaffected by the empty CIPS layers whereas $E_{internal}$ stabilizes the -HP state instead of the -LP state for two or more empty CIPS layers, which effectively doubles the negative polarization value.

The information of $E_{internal}$ and polarization as functions of the number of empty CIPS layers can be used to understand the experimentally observed changes in the polarization hysteresis loops. As an example, we show the polarization for a DC voltage pulse duration of 4 ms and 900 ms as a function of field (instead of voltage) in Figure 4(d). We note that the loops shown in Figure 2(c) are centered at 0 along the y-axis as is standard practice, because current measurements can only to detect a change in polarization during voltage cycling but not their absolute values. However, DFT calculations indicate that only the negative polarization increases, which is the reason for the vertical shift of the hysteresis loop in Figure 4(d) compared to Figure 2(c). Moreover, the loop orientation in Figure 2(c) and Figure 4(d) are opposite due to the definition of the electric field pointing in the direction of the negative potential. First, the $E_{internal}$ is negative, which matches the predicted direction. A negative internal field stabilizes the negative polarization state and needs to be overcome by the externally applied field. Therefore, the polarization loop effectively shifts towards higher positive fields. However, absolute values of the $E_{internal}$ cannot be compared between theory and experiment because the sample thickness is unknown and the formation of conductive empty CIPS layers would effectively change the ferroelectric sample thickness during the experiment. Second, the experimentally observed polarization values are slightly smaller than the theoretically predicted ones. Based on the comparison of theory and experiment, the switchable polarization in the absence of ionic conduction (no empty CIPS layers) is 10.2 µC/cm$^2$ and 6.4 µC/cm$^2$, respectively. In the presence of ionic conduction with three empty CIPS layers, the switchable polarization increases to 14.8 µC/cm$^2$ in theory and 9.8 µC/cm$^2$ in the experiment. These values correspond to 46% and 53% increases, respectively. Therefore, we conclude qualitatively that the stabilization of the -HP state is the origin of the observed polarization increase. The asymmetry between positive and negative polarization values could also explain the difference in switchable polarization for intermediate DC voltage-pulse duration resulting in non-closed hysteresis loops (Figure 2(c)). Closed hysteresis loops are observed (i) after short DC pulses up to 200 ms duration consistent with switching between +LP and -LP states and (ii) after the longest pulse of 900 ms consistent with switching between -HP and +LP states (see also Figure 4). For pulse durations of 400 ms to 800 ms, contributions of behavior (i) and (ii) within different sample areas under the electrode could lead to the observed non-closed hysteresis loops. Moreover, potential imperfections in fitting and separating ferroelectric and ionic contributions to the overall current signal, could contribute to the observed asymmetries in switchable polarization.

## 2.3. Deactivation and Reactivation of Hysteretic Behavior

It is natural to assume that with higher electric fields or longer durations, the Cu ions within CIPS are distributed even more asymmetrically and accumulate at the negative electrode upon ionic current flow. The resulting internal electric field would, therefore, modify coercive voltages more significantly or even lead to complete disorder of Cu in the sample. To explore the impact of activating higher ionic currents, we repeated the measurements on a different capacitor with longer and higher voltage pulses. Initially, we observed a hysteresis loop typical for CIPS (Figure 5(a)) with a polarization around ±5 µC/cm$^2$ and a $V_c^-$ around -3 V and a $V_c^+$ around +5 V. Applying two high voltage pulses of -10 V for 2 and 3 s before the triangular $V$ sweep results in the disappearance of hysteretic behavior (Figure 5(b)). The observed absence of polarization switching is indicative of a strong ionic migration through the sample, which can result in coercive voltages exceeding the applied voltages and loss of polar properties through disordered positions of Cu ions. The absence of leakage current suggests that the applied voltages were not enough to lead to a complete depletion of Cu throughout the sample. It is noteworthy that after a $V$ pulse of +10 V for 1 s, the hysteretic state was restored with polarization and coercive voltage values similar to the initial state (Figure 5(c)). Evidently, deactivation and recovery of ferroelectric switching through high-amplitude voltage pulses provides a path to toggle between hysteretic and non-hysteretic states by modifying the internal ion distribution through high electric fields. The highly reversible nature of Cu redistribution by external fields agrees well with the reversible extraction and reinsertion of Cu from and into CIPS found at higher temperatures.[9] This remarkable reversibility provides exciting opportunities for memory devices beyond binary limitations. In such devices, information could be encoded in three events: (i) switching from positive to negative *P*, (ii) switching from negative to positive *P*, and (iii) no polarization switching within the same voltage range.

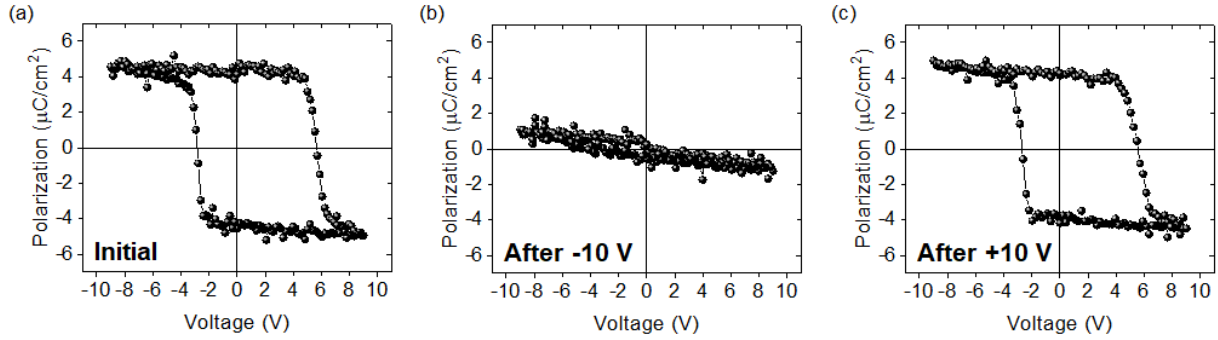

**Figure 5.** Recovery of switching with voltage. Polarization loops of CIPS capacitor (a) initially, (b) after applying a -10 V/ 2+3 s voltage pulses, and (c) after applying a +10 V/ 1 s pulse.

## 3. CONCLUSIONS

We demonstrate ionic control of ferroelectric behavior in $CuInP_2S_6$ capacitors where the position of Cu ions with respect to the layers determines the polarization while simultaneously exhibiting a high mobility. Dependent on the duration of DC voltage pulses, ionic currents become measurable and modify ferroelectric characteristics in subsequent triangular voltage sweeps as measured experimentally. The process of Cu transport is simulated theoretically and the consequences of Cu-free CIPS layers and Cu accumulation under the electrodes on internal electric fields and polarization is explored. Increasing ionic currents, which is equivalent to the formation of an inhomogeneous Cu distribution under the electrodes, lead to internal electric fields that point opposite to the field driving the ionic motion. Therefore, the polarization hysteresis loops become imprinted. In addition, the high- instead of the low-polarization phase is stabilized in the direction of the internal electric field, with the Cu ions displaced slightly into the vdW gaps. This feature doubles the amount of polarization for one polarity, which can explain the ~50% increase in switchable polarization. The interplay of ionic and polar properties can be used to fine-tune ferroelectric characteristics or to fully deactivate and reactivate hysteretic behavior, which provides the exciting opportunity for a three-state ferroelectric capacitor showing either switching between positive and negative polarization states or non-hysteretic characteristics. Moreover, we expect similar phenomena to manifest in other ferroelectric materials where the ionic sublattice that determines polar properties consists of the most mobile ionic species in that compound.

## 4. METHODS

### 4.1 Experimental

CIPS crystal growth was obtained through a solid-state reaction by mixing powders of the four elements and annealing in vacuum at 600 °C for 2 weeks. Scotch tape-based exfoliation was used to transfer flakes onto a 20 nm Ni/90 nm SiO$_2$/Si substrate. Top electrodes were fabricated using electron-beam lithography, electron-beam evaporation, and a lift-off process. The electrode size was 20×20 μm with a thickness of 70 nm (20 nm Ni + 50 nm Au). The capacitor fabrication process is described in detail in Si et al.[16] Three different capacitor sample were used in this work with thicknesses of the CIPS layers of approximately 185 nm (Figure1 and 2), 420 nm (Figure S6 and S7) and 500 nm (Figure5). These values were calculated from the measured coercive voltages and the coercive field of 8.9 mV/nm known from previous measurement on a Ni/CIPS/Ni capacitor sample where the CIPS thickness was known.[15]

The micromanipulator setup consisted of a Cascade Microtech probing station equipped with a Keysight B1500A Semiconductor Device Analyzer and a Radiant Technologies Precision RT66C precision materials analyzer for *I-V* and *P-V* measurements. An Agilent 33220A arbitrary wave generator and a Tektronix TDS5032B oscilloscope are used for the pulsed *I-V* measurements.

The full waveform used for the DC voltage pulse experiment shown in Figure 1 consisted of the square pulse, a first triangular cycle starting with negative voltages and a second triangular cycle. The triangular cycles were applied immediately after the square pulse. The triangular I-V curves in Figure 1 and Figure 2 were acquired at 100 Hz and the polarization loops in Figure 5 were measured at 200 Hz.

### 4.2 Theoretical

The DFT calculations in this study were performed with the Vienna Ab-initio Simulation Package (VASP)[20] where the core-valence-electron interactions are described via the projected-augmented-wave (PAW) method.[21-22] The wave functions were expanded in a plane-wave basis using a 400 eV energy cutoff. Exchange-correlation effects were described using the Perdew-Burke-Ernzerhof[23] generalized gradient approximation including van der Waals corrections via Grimme's DFT-D3 method with Becke-Johnson (BJ) damping.[24-25] The system was modeled using a supercell containing 6 layers of (2 × 1) CuInP$_2$S$_6$ and 3 layers of (4 × 3$\sqrt{2}$) Ni(111) on both sides of CIPS with a vacuum layer of 15 Å. Spin polarization of the charge density was included in the calculations due to the high Curie temperature of Ni. During structural optimization, the Brillouin zone was sampled using a Γ-centered 4 × 4 × 1 k-point mesh and all atoms were fully relaxed until the net forces on each were smaller than 0.01 eV/Å. For density of states (DOS) calculations, the k-point sampling was increased to a denser 6 × 6 × 1 grid. The DOS plots were generated using Gaussian integration with broadening factor of $\sigma = 0.1$ eV. For determination of the extracted band edges, the Brillouin zone was integrated using the tetrahedron method[26] which gives sharper band edges.

**SUPPORTING INFORMATION**

Details on measured current at the start and end of DC pulses, current during and after positive DC pulses, details on fitting to separate ionic from ferroelectric current contributions, detailed analysis of ionic current/charge, polarization hysteresis area, study on impact of cycle number on coercive fields, details on DOS calculations and DFT of electric field and polarization

**ACKNOWLEDGEMENTS**

Experiments and analysis were supported by the U.S. Department of Energy (DOE), Office of Science, Basic Energy Sciences, Materials Sciences and Engineering Division. Part of the analysis and manuscript writing was supported by the Center for Nanophase Material Sciences, which is a U.S. DOE Office of Science User Facility. Theory and analysis, performed at Vanderbilt University, was supported by the U.S. DOE, Office of Science, Basic Energy Sciences, Materials Sciences and Engineering Division grant No DE-FG02-09ER46554 and the McMinn Endowment. Computations were performed using resources provided by the National Energy Research Scientific Computing Center (NERSC), a U.S. Department of Energy Office of Science User Facility operated under Contract No. DE-AC02-05CH11231 and by the Department of Defense's High-Performance Computing Modernization Program (HPCMP).

**Abstract Figure:**

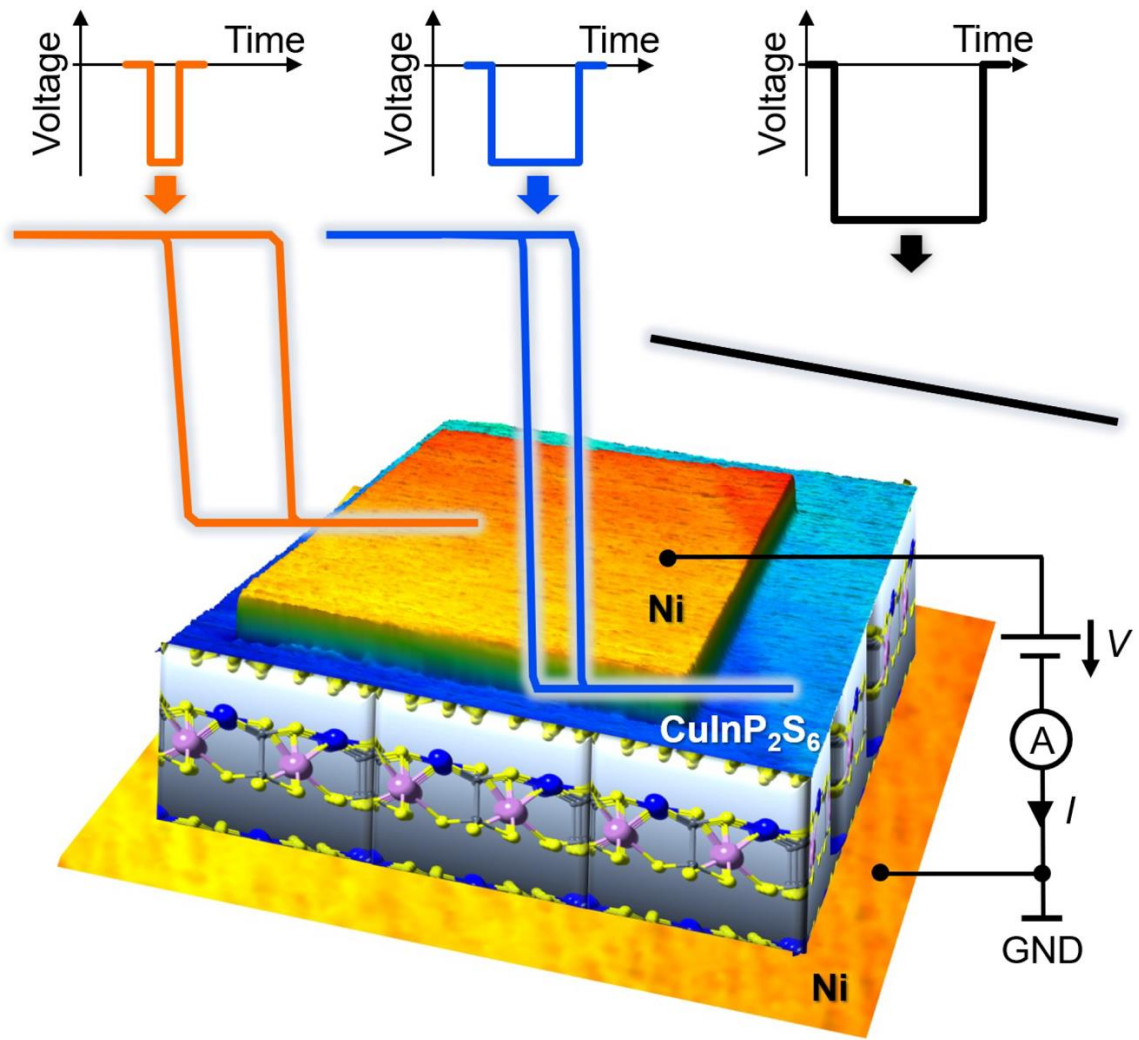